\documentclass[journal]{IEEEtran}
\ifCLASSINFOpdf
\else
\fi
\usepackage{subfigure,cite,graphicx,amsmath,amssymb,mathrsfs,epsf}
\usepackage[usenames]{color}
\usepackage{multirow}
\usepackage{tabularx}
\usepackage{stfloats}
\usepackage{setspace}
\usepackage{algpseudocode,algorithmicx,tabularx}
\usepackage{algorithm}

\makeatletter
\newcommand{\multiline}[1]{%
  \begin{tabularx}{\dimexpr\linewidth-\ALG@thistlm}[t]{@{}X@{}}
    #1
  \end{tabularx}
}
\makeatother

\begin{document}
%
\title{Transport Capacity Optimization for Resource Allocation in Tera-IoT Networks}
%
%
%

\author{Cheol~Jeong,~\IEEEmembership{Member,~IEEE},
        Chang-Jae~Chun,~\IEEEmembership{Member,~IEEE},
        Won-Yong~Shin,~\IEEEmembership{Senior Member,~IEEE},\\
        and Il-Min Kim,~\IEEEmembership{Senior Member,~IEEE}
\thanks{The work of C. Jeong was supported by the National Research Foundation of Korea (NRF) grant funded by the Korea government (MSIT) (No. 2020R1F1A1076721). The work of W.-Y. Shin was supported by NRF grant funded by the Korea government (MSIT) (No. 2021R1A2C3004345) and by the Yonsei University Research Fund of 2021 (2021-22-0083). The work of I.-M. Kim was supported by the National Sciences and Engineering Research Council of Canada (NSERC).}
\thanks{C.~Jeong is with the Department of Intelligent Mechatronics Engineering and the Department of Convergence Engineering for Intelligent Drone, Sejong University, Seoul 05006, Republic of Korea. E-mail: cheol.jeong@ieee.org.}
\thanks{C.-J. Chun is with the Department of Artificial Intelligence, Sejong University, Seoul 05006, Republic of Korea. E-mail: cjchun84@gmail.com.}%
\thanks{W.-Y.~Shin (corresponding author) is with the School of Mathematics and Computing (Computational Science and Engineering), Yonsei University,
Seoul 03722, Republic of Korea. E-mail: wy.shin@yonsei.ac.kr.}
\thanks{I.-M.~Kim is with the Department of Electrical and Computer Engineering, Queen's University, Kingston, ON K7L 3N6, Canada. E-mail: ilmin.kim@queensu.ca.}
%
}

\newtheorem{definition}{Definition}
\newtheorem{theorem}{Theorem}
\newtheorem{lemma}{Lemma}
\newtheorem{example}{Example}
\newtheorem{corollary}{Corollary}
\newtheorem{proposition}{Proposition}
\newtheorem{conjecture}{Conjecture}
\newtheorem{remark}{Remark}

\def \diag{\operatornamewithlimits{diag}}
\def \min{\operatornamewithlimits{min}}
\def \max{\operatornamewithlimits{max}}
\def \log{\operatorname{log}}
\def \max{\operatorname{max}}
\def \rank{\operatorname{rank}}
\def \out{\operatorname{out}}
\def \exp{\operatorname{exp}}
\def \arg{\operatorname{arg}}
\def \E{\operatorname{E}}
\def \tr{\operatorname{tr}}
\def \SNR{\operatorname{SNR}}
\def \dB{\operatorname{dB}}
\def \ln{\operatorname{ln}}

\def \bmat{ \begin{bmatrix} }
\def \emat{ \end{bmatrix} }

\def \be {\begin{eqnarray}}
\def \ee {\end{eqnarray}}
\def \ben {\begin{eqnarray*}}
\def \een {\end{eqnarray*}}

\maketitle

%
\IEEEpeerreviewmaketitle

%
%
%
%



\begin{abstract}
We present a new adaptive resource optimization strategy that jointly allocates the subwindow and transmit power in multi-device terahertz (THz) band Internet of Things (Tera-IoT) networks. Unlike the prior studies focusing mostly on maximizing the sum distance, we incorporate {\em both rate and transmission distance} into the objective function of our problem formulation with key features of THz bands, including the spreading and molecular absorption losses. More specifically, as a performance metric of Tera-IoT networks, we adopt the {\em transport capacity (TC)}, which is defined as the sum of the {\em rate--distance products} over all users. This metric has been widely adopted in large-scale ad hoc networks, and would also be appropriate for evaluating the performance of various Tera-IoT applications. We then formulate an optimization problem that aims at maximizing the TC. Moreover, motivated by the importance of the transmission distance that is very limited due to the high path loss in THz bands, our optimization problem is extended to the case of allocating the subwindow, transmit power, and transmission distance. We show how to solve our problems via an effective two-stage resource allocation strategy. We demonstrate the superiority of our adaptive solution over benchmark methods via intensive numerical evaluations for various environmental setups of large-scale Tera-IoT networks.
\end{abstract}


%
\IEEEpeerreviewmaketitle

\section{Introduction}~\label{SEC:Introduction}

\subsection{Background}

Terahertz (THz) band (0.1--10 THz~\cite{ITU-R-2352}) communications have been envisioned as a highly promising technology to overcome the scarcity of spectrum resources in current wireless systems~\cite{AkyldizJornetHan:PC14,GhafoorBoujnahRehmaniDavy:20}. Promising applications include nanoscale applications such as ultra-dense Internet of Nano-Things (IoNT)~\cite{AkyildizJornet:10,AkkariWangJornetFadelElrefaeiMalikAlmasriAkyildiz:IoT16}, plant monitoring nanosensor networks~\cite{AfsharinejadDavyJenningsBrennan:IoT16}, and wireless body area networks. The Internet of Things (IoT) is evolving toward holographic communications~\cite{GuoYuZhangLiJiLeung:IoT21} and immersive virtual reality (VR)~\cite{DuYuLuWangJiangChu:IoT20,FantacciPicano:IoT,ChaccourSoorkiSaadBennisPopovski:arXiv20,ChaccourSoorkiSaadBennisPopovski:ICC20} using a very high data rate at THz bands with a short distance. Although the millimeter wave (mmWave) band can also provide a high data rate, it cannot accommodate multiple users who request high rates simultaneously. Thus, the THz bands have received substantial attention very recently for 6G wireless communications~\cite{GuiLiuTangKatoAdachi:20}. Despite such huge demands, designing ready-to-use THz communication systems leads to new research challenges that have never been encountered by any existing communication systems operating at lower frequencies (e.g., the mmWave spectrum). This is due to the fact that the THz signals suffer from the inevitable high path loss, which is mainly induced by both the spreading effect during propagation and the absorption effect such as molecular absorption~\cite{AkyldizJornetHan:PC14,SongNagatsuma:TTST11}.

On the other hand, while the problem of resource allocation has attracted a great deal of attention in wireless systems operating at {\em lower frequencies} over several decades~\cite{SadrAnpalaganRaahemifar:CST09,XuGuiGacaninAdachi:21}, the developed solutions cannot be straightforwardly applied to the THz band since they did not take into account the distinguishable THz channel properties such as high frequency selectivity caused by irregular absorption profiles. For the THz band, research on the resource allocation has been largely underexplored.

\subsection{Main Contributions}

In this paper, we study a multi-carrier {\em THz band IoT (Tera-IoT)} network deploying a number of {\em small-scale} devices (receivers). In our Tera-IoT network model, we introduce a new adaptive resource allocation strategy by performing joint allocation of the subwindow and transmit power.

The transmission rate has been a long-standing key performance indicator (KPI) of wireless systems. Thus, resource allocation 
has primarily been carried out by aiming at maximizing the sum rate. In the THz bands, however, transmission rate alone might not be a good KPI since a very high data rate can readily be achieved through the abundant bandwidth available in THz bands. Instead, the transmission distance can be a rather reasonable and convincing KPI of THz communications, as already stated in several studies~\cite{HanBicenAkyildiz:TSP16,HanAkyildiz:TTST16}. When it comes to two KPIs including the rate and the distance, an important question arising is: ``What is the most appropriate performance metric which well represents the performance of {\em Tera-IoT} networks?" Let us consider two specific scenarios: 1) a device can be served by an access point (AP) (transmitter) at a very high data rate but the transmission distance is very short and 2) a device can be served by the AP that is far apart but at a very low data rate. Between these scenarios, one cannot easily determine which scenario will be more suitable for improving the performance of multi-device Tera-IoT networks. In this paper, we claim that what really matters for the network performance is both how fast and how far bits are transmitted. Therefore, different from the prior studies in~\cite{HanBicenAkyildiz:TSP16,HanAkyildiz:TTST16,LinLi:TC15,ZhangHanWang:SECON19,ZhangZhangLiuLongDongLeung:JSAC20,ZhangDuanLongLeung:TC,ZhangWangPoor:JSAC21,ZhaiTangPengWang:JSAC21}, we are interested in {\em not only the rate but also the distance} with which bits are transmitted.

Mathematically, we jointly incorporate both the rate and the distance into the objective function of our problem formulation by considering the most peculiar features of THz bands, including the spreading and molecular absorption losses.
To this end, instead of solely maximizing either the sum rate or the sum distance, we aim at formulating a {\em transport capacity (TC)} maximization problem. The TC~\cite{GuptaKumar:00,XueXieKumar:05,JovicicViswanathKulkarni:04,AndrewsWeberKountourisHaenggi:10}, defined as the sum of the {\em rate--distance products} over all devices, has been widely adopted as a fundamental performance measure to analyze the scaling behavior of ad hoc {\em dense} networks having a large number of users within a fixed area. In the same context, the TC must be a suitable metric in evaluating the performance of Tera-IoT dense networks along with the THz channel properties since 1) many THz applications would fit into dense network environments deploying a number of devices, e.g., densely connected IoNT~\cite{PetrovMoltchanovKoucheryavy:15}, and 2) the TC well characterizes the fundamental limits of multiuser networks as long as path loss or other attenuation models are concerned~(refer to~\cite{Peel:12} for more details).

Adopting the TC as a KPI, we present two optimization problems and their solutions. We first formulate the TC maximization problem when the transmission distance between the AP and each device is given and fixed. We then study how to effectively solve the joint subwindow and power optimization problem by presenting a {\em two-stage} strategy, which is challenging since our formulation based on the TC is analytically intractable. 
Specifically, in the first stage, we carry out the subwindow assignment using the Hungarian method given a transmit power. In the second stage, for a given subwindow assignment, we allocate the transmit power to each device. Additionally, motivated by the importance of the transmission distance that is very limited due to the high path loss in THz bands, we extend our optimization problem to the case where the transmission distance between the AP and each device is allowed to vary to potentially improve the performance. To solve this problem, we jointly allocate the subwindow, transmit power, and transmission distance in the sense of maximizing the TC under {\em heterogeneous rate} constraints. The extended problem can be solved according to our two-stage strategy; in the second stage, the transmit power and distance are jointly optimized in an iterative manner by discovering a relationship between the optimal distance and power.

To validate the effectiveness of the proposed resource allocation method, we perform intensive numerical evaluations for various environmental setups of multi-device Tera-IoT networks. For the first scenario where the transmission distance between the AP and each device is fixed, we show the benefits of our TC maximization method over the sum rate maximization method in terms of device fairness. For the second scenario where the transmission distance varies, we demonstrate that our adaptive solution to the joint subwindow, power, and distance optimization problem consistently outperforms benchmark methods in terms of the TC. Interestingly, we show that 1) under the same rate constraint for each device, the gain of our adaptive method over the distance maximization method tends to be large when the required rate is set low, 2) our method yields a significant improvement when our problem is formulated with heterogeneous rate requirements, 3) the allocated distance depends dominantly on the absorption loss as well as the frequency, 4) there is a fundamental trade-off between the rate and the transmission distance, and last but not least 5) the rate--distance trade-off can be improved by the proposed adaptive method.

\subsection{Organization}
The rest of this paper is organized as follows. In Section~\ref{SEC:RelatedWork}, we summarize studies that are related to our work. In Section~\ref{SEC:ChannelSystem}, the channel and network models are described. In Section~\ref{SEC:OptimizationFixedDistances}, our TC maximization problem with fixed distances is formulated and solved. In Section~\ref{SEC:Optimization}, the optimization problem is further extended by including the distance optimization and the effective solution to our problem is derived. In Section~\ref{SEC:NumericalResults}, numerical evaluations are performed to validate our method. Finally, Section~\ref{SEC:Conclusion} summarizes the paper with some concluding remarks.

\section{Related Work}~\label{SEC:RelatedWork}

The method that we propose in this paper is related to two topics, namely the physical layer design of THz communications and the resource allocation in THz bands.

{\bf Physical layer design.} There has been a steady push to develop ultra-high-speed communication methods in the THz band from the physical layer perspective such as 1) THz channel modeling~\cite{PiesiewiczJansenMittlemanOstmannKochKurner:TAP07,PriebeKurner:TWC13,JornetAkyildiz:TWC11,HanBicenAkyildiz:TWC15,HanChen:CM18}, 2) waveform design along with adaptation of both the transmit power and the number of frames~\cite{HanBicenAkyildiz:TSP16}, 3) hybrid analog and digital beamforming~\cite{LinLi:TWC15,LinLi:CM16,YuanYangYangHanAn:TC20,YanHanYuan:JSAC20,HuangYang:JSAC21,ZhangHaoSunYang:ISJ}, and 4) network massive multiple-input multiple-output (MIMO)~\cite{YouChenSongJiangWangGaoFettweis:JSAC20}. Although there are hardware challenges on implementing orthogonal frequency division multiplexing (OFDM) in THz bands~\cite{HanAkyildiz:TTST16} such as the high peak-to-average power ratio (PAPR) and strict requirements for frequency synchronization, there have been lab experiments attempting to realize OFDM systems in THz bands~\cite{HermeloShihSteegNgomaStohr:17,SenJornet:19}. In~\cite{HermeloShihSteegNgomaStohr:17}, the wireless THz transmission around 325 GHz was experimentally demonstrated using 64-quadrature amplitude modulation (QAM) in an OFDM system. In~\cite{SenJornet:19}, at about 1.02 THz, tens of Gbps transmission over sub-meter distances was tested using OFDM. In~\cite{YouChenSongJiangWangGaoFettweis:JSAC20,ZhangHaoSunYang:ISJ}, OFDM systems were adopted in their models to effectively use THz bands.

{\bf Resource allocation.} Subcarrier assignment, bit loading, and transmit power allocation methods in wireless systems operating at {\em lower frequencies} were studied for multiuser OFDM systems (e.g.,~\cite{YinLiu:00,WongTsuiChengLetaief:99,LeTranVuJayalath:TC05}). However, for the THz band, research on the resource allocation and scheduling has been largely underexplored except for only a few attempts~\cite{HanAkyildiz:TTST16,LinLi:TC15,ZhangHanWang:SECON19,ZhangZhangLiuLongDongLeung:JSAC20,ZhangDuanLongLeung:TC,ZhangWangPoor:JSAC21,ZhaiTangPengWang:JSAC21}. As the state-of-the-art method of resource allocation in THz communications, a joint optimization framework for performing subwindow allocation, modulation adaptation, and transmit power control was presented in~\cite{HanAkyildiz:TTST16} in the sense of maximizing the {\em sum transmission distance} under the minimum required rate constraints for multi-carrier THz systems. Unlike the prior studies on the resource allocation at lower frequencies, the optimization framework in~\cite{HanAkyildiz:TTST16} was designed by exploiting the unique THz channel properties such as the distance--frequency dependence. More precisely, distance-aware bandwidth-adaptive spectrum allocation was performed based on a principle that a long-distance user uses a subwindow at the central window while a short-distance user uses a subwindow at the edge of window, where the window is a band of frequencies at the valley between two high peak path losses on the path loss profiles of THz bands. In~\cite{LinLi:TC15}, an adaptive power allocation and antenna subarray selection strategy was also developed to minimize the number of subarrays under a minimum rate requirement for multiuser THz systems equipped with multiple antenna subarrays at the transmitter. More flexibility in the subwindow assignment was provided by allowing a short-distance user to use a subwindow at the central window as well as at the edge of window. In~\cite{ZhangHanWang:SECON19}, following the spectrum allocation principle in~\cite{HanAkyildiz:TTST16}, beamforming design and power-bandwidth allocation were presented in THz non-orthogonal multiple access (NOMA) systems. In~\cite{ZhangZhangLiuLongDongLeung:JSAC20,ZhangDuanLongLeung:TC}, the energy efficiency maximization was studied in NOMA systems. Specifically, user scheduling and power allocation in~\cite{ZhangZhangLiuLongDongLeung:JSAC20} and subchannel assignment and power allocation in~\cite{ZhangDuanLongLeung:TC} were conducted. As the limited energy at THz-band devices is a big challenge, the capacity was maximized by adopting energy harvesting techniques in~\cite{ZhangWangPoor:JSAC21}. In~\cite{ZhaiTangPengWang:JSAC21}, in order to serve users dispersed in a large angular range, user grouping and beam spreading were proposed. Moreover, research attention has been paid to resource allocation in unmanned aerial vehicle (UAV)-assisted THz networks~\cite{Xu:arXiv20,Pan:WCL21}. In~\cite{Xu:arXiv20}, the deployment of UAVs serving ground users was optimized in the THz band. In~\cite{Pan:WCL21}, the UAV's trajectory, subwindow, and power were jointly optimized.

{\bf Discussions.} The aforementioned studies in~\cite{HanAkyildiz:TTST16,LinLi:TC15,ZhangHanWang:SECON19,ZhangZhangLiuLongDongLeung:JSAC20,ZhangDuanLongLeung:TC} on the resource allocation aimed at discovering effective transmission policies by either enhancing the transmission distance or restricting the number of subarrays with limited radio frequency (RF) chains that are used to compensate for the high attenuation of THz signals. Although such studies offer meaningful solutions for THz communication systems, it still remains an open challenge how to improve the {\em system-wise performance} given {\em severely limited} resources.

\section{THz Band Channel and Network Models}~\label{SEC:ChannelSystem}
In this section, we describe the characteristics of the THz band channel and our multi-device multi-carrier Tera-IoT network.

\subsection{THz Band Channel Model}
\begin{figure}[!t]
  \centering
  \includegraphics[height=.37\textwidth]{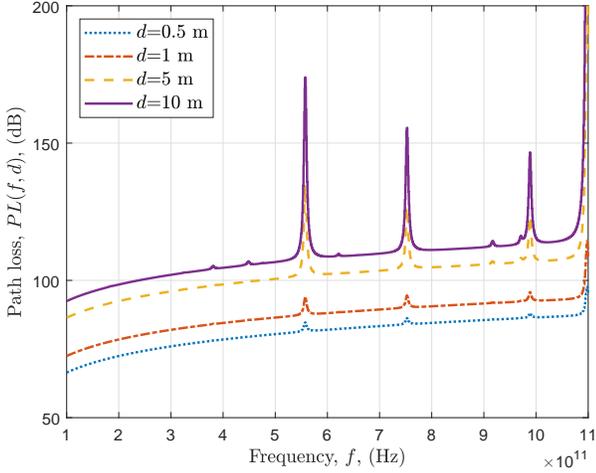}
  \caption{The path loss in the THz band for different transmission distances.}\label{Fig:Pathloss-frequency}
\end{figure}

There have been a lot of studies on characterizing the peculiarities of the THz band~\cite{PiesiewiczJansenMittlemanOstmannKochKurner:TAP07,PriebeKurner:TWC13,JornetAkyildiz:TWC11,LinLi:TC15}. In particular, to model indoor THz channels, a deterministic model built upon reflection coefficients for common indoor building materials and a stochastic spatio--temporal model were presented in~\cite{PiesiewiczJansenMittlemanOstmannKochKurner:TAP07} and~\cite{PriebeKurner:TWC13}, respectively. Moreover, in~\cite{JornetAkyildiz:TWC11}, a general line of sight (LOS) channel model that covers from 0.1 THz to 10.0 THz was developed by using the radiative transfer theory~\cite{GoodyYung:89} and the information in the HITRAN database~\cite{Rothman:09}. The multi-path THz channel model was developed based on ray tracing techniques~\cite{HanBicenAkyildiz:TWC15}. Among those, we adopt the channel model in~\cite{JornetAkyildiz:TWC11}, which properly characterizes the THz band regime of interest alongside analytical tractability. According to the model in~\cite{JornetAkyildiz:TWC11}, the long-term channel power gain, denoted by $|h(f,d)|^2$, consisting of the spreading loss $L_{\textrm{spread}}(f,d)$ and the molecular absorption loss $L_{\textrm{abs}}(f,d)$, is expressed as 
\begin{align}\label{THz-ChannelGain}
  |h(f,d)|^2 = G_t G_r L_{\textrm{spread}}(f,d) L_{\textrm{abs}}(f,d),
\end{align}
where $G_t$ and $G_r$ are the transmitter and receiver antenna gains, respectively; $f$ is the carrier frequency; and $d$ is the distance between a transmitter and a receiver. Here, the spreading loss $L_{\textrm{spread}}(f,d)$ accounts for the power spread as a wave propagates through the medium and is defined as
\begin{align}
  L_{\textrm{spread}}(f,d):=\left( \frac{c}{4\pi fd}\right)^2 ,
\end{align}
where $c$ is the speed of light in free space. The molecular absorption loss $L_{\textrm{abs}}(f,d)$ represents a loss caused by the absorption from molecules such as water vapor in the atmosphere and is given by
\begin{align}
  L_{\textrm{abs}}(f,d) := e^{-K_{\textrm{abs}}(f)d},
\end{align}
where $K_{\textrm{abs}}(f)$ is the absorption coefficient at frequency $f$. The path loss (in dB) is given by 
\begin{align}
    PL(f,d)=-10\log_{10} |h(f,d)|^2.
\end{align}
Similarly as in Fig. 2 in~\cite{LinLi:TC15}, Fig.~\ref{Fig:Pathloss-frequency} plots the path loss versus different transmission distances of $d=0.5,1,5$, and $10$ m using the reference code and the absorption coefficients presented in~\cite{HossainXiaJornet:19}.\footnote{The discrepancy between this figure and Fig. 2 in~\cite{LinLi:TC15} comes from the fact that the data of the absorption coefficients in~\cite{HossainXiaJornet:19} may be coarse in terms of frequency.} It is observed that there are very high peaks around 550 GHz, 750 GHz, 1 THz, and 1.1 THz especially when the transmission distance $d$ is long.

\subsection{Network Model}
We consider a multi-device multi-carrier Tera-IoT network, where an AP is deployed to serve $K$ (selected) IoT devices. Similarly as in~\cite{YouChenSongJiangWangGaoFettweis:JSAC20, ZhangHaoSunYang:ISJ}, OFDM is adopted with $N$ subwindows, each of which occupies the bandwidth $W$ Hz. We assume that there does not exist the inter-band interference from neighboring subwindows nor the inter-symbol interference since such interference is typically neglected in OFDM systems. In our study, we only take into account the case of $K\leq N$ by assuming that those $K$ devices are scheduled among all possible devices by medium access control (MAC) protocols~\cite{GhafoorBoujnahRehmaniDavy:20} if the total number of devices is greater than $N$. We assume that each device uses only one subwindow and each subwindow can be used by only one device. It is noteworthy that a low-cost device is possible if only a single subwindow per device is supported and the bandwidth of a subwindow can be set sufficiently large so as to provide multi-Gbps since the coherence bandwidth in THz band is large~\cite{HanChen:CM18}. We focus on the use of analog beamforming that can support $K$ analog beams at the AP (transmitter) in order to serve $K$ devices in different locations, which is preferred for device-specific data transmission at high frequencies (refer to~\cite{MolischRatnamHanLiNguyenLiHaneda:ComMag2017} and references therein). The transmission rate $R_{k,n}(d_k)$ of the $k$th device on the $n$th subwindow at distance $d_k$ can be written as
\begin{align}\label{Rate}
  R_{k,n}(d_k)=W\log_2\left(1+\frac{p_{k}|h(f_n,d_k)|^2}{\sigma^2}\right),
\end{align}
where $p_k$ is the power allocated to the $k$th device; $f_n$ is the frequency of the $n$th subwindow; $d_k$ is a distance between the AP and the $k$th device; $\sigma^2=N_0 W$ is the noise power per subwindow; and $N_0$ is the power spectral density of additive white Gaussian noise (AWGN).
By substituting the channel power gain of the THz band in~\eqref{THz-ChannelGain} into~\eqref{Rate}, the transmission rate of the $k$th device on the $n$th subwindow at distance $d_k$ can be rewritten as
\begin{align}\label{Rate2}
  R_{k,n}(d_k) \!= \!W\log_2 \left(1+ \frac{p_k G_t G_r}{\sigma^2}e^{-K_{\textrm{abs}}(f_n)d_k} \left( \frac{c}{4\pi f_n d_k}   \right)^2 \right).
\end{align}
Then, the transmission rate of the $k$th device over all subwindows at distance $d_k$ is given by
\begin{align}\label{RatePerDevice}
    R_k(d_k) = \sum_{n=1}^N \rho_{k,n} R_{k,n}(d_k),
\end{align}
where $\rho_{k,n}\in \{0,1\}$ is the indicating bit such that $\rho_{k,n}=1$ only if the $n$th subwindow is allocated to the $k$th device and it is set to zero otherwise.\footnote{To simplify notations, $R_k(d_k)$ will be written as $R_k$ when dropping $d_k$ does not cause any confusion.} 

\subsection{Transport Capacity}
In the THz band, a ultra-high transmission rate, up to several Gbps, can easily be achieved by virtue of a very wide bandwidth. Due to the very high path loss, however, the transmission distance is severely limited in THz communications. Thus, the transmission distance can be regarded as one of crucial KPIs of THz communication systems while the data rate, area traffic capacity, latency, and so forth have been adopted as KPIs of the standard fifth generation (5G) systems. Nevertheless, there is still an open challenge on discovering a better KPI suited for evaluating the performance of THz networks since traditional optimization such as the rate maximization does not necessarily maximize the network performance that can be represented by both {\em how fast} and {\em how far} the bits are transmitted.

In our study, rather than solely maximizing the rate, we present a new resource allocation approach for taking into account both rate and distance simultaneously by adopting the {\em TC}~\cite{GuptaKumar:00,XueXieKumar:05,JovicicViswanathKulkarni:04,AndrewsWeberKountourisHaenggi:10} as our objective function, which is defined as the sum of the {\em rate--distance products} over all devices. When we denote $T_k$ as the rate--distance product of device $k$, from (\ref{Rate2}) and (\ref{RatePerDevice}), the TC (in $m\cdot\text{bps}$), $T$, is given by
\begin{align} \label{TC}
    T &=\sum_{k=1}^K T_k  \nonumber\\
    &= \sum_{k=1}^K  d_k R_k \nonumber\\
    &=\sum_{k=1}^K \sum_{n=1}^N \rho_{k,n}  d_k W
    \nonumber\\
    &~~~~~~~~~~~\cdot\log_2 \left(1+\frac{p_k G_t G_r}{\sigma^2}e^{-K_{\textrm{abs}}(f_n)d_k}\left( \frac{c}{4\pi f_n d_k}   \right)^2 \right).
\end{align}

\section{TC Maximization with Fixed Distances}\label{SEC:OptimizationFixedDistances}

In this section, we formulate the TC maximization problem in our multi-device Tera-IoT network when the distance from the AP (transmitter) to each device is {\em given and fixed}. Our solution to the problem consists of two stages: 1) subwindow assignment and 2) power allocation.

\subsection{Problem Formulation}
We will formulate a problem in order to maximize the TC when the distance between each device and the AP is given and fixed. More specifically, we aim at allocating subwindows to the devices and optimizing the transmit power while satisfying the transmit power constraints according to the following constrained optimization:
\begin{subequations} \label{Main-Opt-Fixed}
\begin{align}
\underset{\{\rho_{k,n}\},\{p_k\}}\max&~~ T
\label{fix1}\\
\textrm{s.t.}~~~~~~~&~~ \rho_{k,n} \in \{0,1\},~~~\forall k,n \label{Main-Opt-Fixed-Const-1} \\
&~~ \sum_{k=1}^K \rho_{k,n} \leq 1,~~~\forall n \label{Main-Opt-Fixed-Const-2} \\
&~~ \sum_{n=1}^N \rho_{k,n} \leq 1,~~~\forall k \label{Main-Opt-Fixed-Const-3} \\
&~~ p_k \geq 0,~~~\forall k  \label{Main-Opt-Fixed-Const-4}\\
&~~ \sum_{k=1}^K  p_k \leq P_{\textrm{T}}, \label{Main-Opt-Fixed-Const-5}
\end{align}
\end{subequations}
where $T$ is the TC in (\ref{TC}) and $P_{\textrm{T}}$ is the maximum total transmit power. Here, \eqref{Main-Opt-Fixed-Const-2} and \eqref{Main-Opt-Fixed-Const-3} are needed to ensure that a subwindow can be allocated to at most one device and each device can utilize only one subwindow. The constraints in \eqref{Main-Opt-Fixed-Const-5} is the total transmit power constraint. We note that the optimization problem in \eqref{Main-Opt-Fixed} is nonconvex due to the integer constraints.

In general, it is quite difficult to solve such a nonconvex optimization problem with integer constraints. In order to tackle the nonconvex mixed-integer nonlinear programming in~\eqref{Main-Opt-Fixed}, we propose a {\em two-stage} approach in which the subwindows are assigned and the transmit power is then allocated. In the subwindow assignment, it is assumed that $p_k$ is set to $P_{\textrm{T}}/K$, i.e., equal power allocation. The TC is maximized using the Hungarian method~\cite{Kuhn:55}. In the power allocation, given the subwindow assignment, the values of the transmit power are determined so as to maximize the TC. In the following subsections, we present the proposed two-stage iterative allocation strategy precisely.

\subsection{Subwindow Assignment} \label{SEC:Subwindow_fixed}

In this subsection, we describe the first stage of the nonconvex mixed-integer nonlinear programming, which corresponds to the subwindow assignment. In this stage, assuming that the values of $p_k=P_{\textrm{T}}/K$, we determine optimal $\rho_{k,n}$ in terms of maximizing the TC. The subwindow assignment problem can be written as

\begin{align}\label{Prob-Subwindow-Fixed}
\underset{\{ \rho_{k,n}\}}\max &~~ \sum_{k=1}^K \sum_{n=1}^N \rho_{k,n}d_k R_{k,n}(d_k)\nonumber\\
\textrm{s.t.}~&~~ \rho_{k,n} \in \{0,1\},~~~\forall k,n \nonumber\\
&~~ \sum_{k=1}^K \rho_{k,n} \leq 1,~~~\forall n \nonumber\\
&~~ \sum_{n=1}^N \rho_{k,n} \leq 1,~~~\forall k.
\end{align}
The optimal solution to the above problem can always be found using the Hungarian method~\cite{Kuhn:55}, which is a combinatorial optimization algorithm and has been widely used in the resource allocation for multi-carrier communication systems~\cite{WongTsuiChengLetaief:99,YinLiu:00,HanJiLiu:TC05,LiuDai:TSP14}. After the subwindow assignment, the subwindow index that has been allocated to the $k$th device is denoted by $n_k$.

\subsection{Power Allocation}

In this subsection, we describe the second stage of our allocation strategy. More concretely, given the subwindow assignment, we optimally allocate the transmit power to maximize the TC under the total power constraint. The power allocation problem is expressed as
\begin{align}\label{Main-Opt-Fixed-Power}
\underset{\{p_k\}}\max& \sum_{k=1}^K d_k W\log_2 \left(1+ \frac{p_k G_t G_r}{\sigma^2}e^{-K_{\textrm{abs}}(f_{n_k})d_k} \left( \frac{c}{4\pi f_{n_k} d_k}   \right)^2 \right) \nonumber\\
\textrm{s.t.}
&~~ p_k \geq 0,~~~\forall k  \nonumber\\
&~~ \sum_{k=1}^K  p_k \leq P_{\textrm{T}}.
\end{align}

The optimization problem in~\eqref{Main-Opt-Fixed-Power} can be regarded as a weighted sum rate maximization where the distance from the AP to the $k$th device, $d_k$, corresponds to the weight for the transmission rate of the $k$th device. In~\cite{HooHalderTelladoCioffi:TC04}, it was shown that multilevel water-filling is the optimal solution to the weighted sum rate maximization. Similarly, the optimal solution to the problem in~\eqref{Main-Opt-Fixed-Power} is given by
\begin{align}
   p_k = \left[\frac{d_k}{\lambda} - \frac{\sigma^2}{G_t G_r e^{-K_{\textrm{abs}}(f_{n_k})d_k}}\left(\frac{4\pi f_{n_k}d_k}{c}\right)^2\right]^+,~~~\forall k,
\end{align}
where $\lambda$ is determined such that the total power constraint in~\eqref{Main-Opt-Fixed-Power} is fulfilled with an equality.

\section{TC Maximization with Variable Distances}~\label{SEC:Optimization}

In this section, we extend our TC maximization problem in~\eqref{Main-Opt-Fixed} by allowing the distance between the AP and each device to vary in order to further improve the performance. Then, we elaborate on our two-stage resource allocation strategy that effectively assigns the subwindow, power, and distance.

\subsection{Problem Formulation}

In the THz band, while a ultra-high transmission rate can be readily achieved by virtue of a very wide bandwidth, a very long transmission distance cannot be guaranteed easily due to the very high path loss. In this context, improving the coverage area is more important in the THz band than the case of lower frequencies (e.g., the legacy sub-6 GHz band). To this end, the distance optimization problem in the THz band was studied for diverse applications (refer to\cite{HanBicenAkyildiz:TSP16,HanAkyildiz:TTST16,Xu:arXiv20} and references therein).

In practice, there are different needs of devices along with {\em heterogeneous} transmission rates. In this section, we formulate a TC maximization problem under heterogeneous rate constraints in which each device has a different rate requirement. More specifically, we determine the subwindow, the transmit power, and the distance for each device in the sense of maximizing the TC while satisfying the transmit power constraints and the minimum rate constraints. The constrained optimization problem can be formulated as
\begin{subequations} \label{Main-Opt}
\begin{align}
\underset{\{\rho_{k,n}\},\{p_k\}, \{d_k\}}\max&~~ T
\label{fix1}\\
\textrm{s.t.}~~~~~~~&~~ \rho_{k,n} \in \{0,1\},~~~\forall k,n \label{Main-Opt-Const-1} \\
&~~ \sum_{k=1}^K \rho_{k,n} \leq 1,~~~\forall n \label{Main-Opt-Const-2} \\
&~~ \sum_{n=1}^N \rho_{k,n} = 1,~~~\forall k \label{Main-Opt-Const-3} \\
&~~ p_k \geq 0,~~~\forall k  \label{Main-Opt-Const-4a}\\
&~~ d_k \geq 0,~~~\forall k  \label{Main-Opt-Const-4b}\\
&~~ \sum_{k=1}^K  p_k \leq P_{\textrm{T}} \label{Main-Opt-Const-5} \\
&~~ R_k \geq R_{k,\textrm{th}},~~~\forall k, \label{Main-Opt-Const-6}
\end{align}
\end{subequations}
where the constraints in~\eqref{Main-Opt-Const-6} are the minimum rate constraints. It is worth noting that the rate constraints $R_{k,\textrm{th}}$ can be set differently according to the target applications or services of devices. Since the objective function is nonconvex in $d_k$ and the integer constraints are also nonconvex, the optimization problem in \eqref{Main-Opt} is nonconvex.

As in~\eqref{Main-Opt-Fixed}, it is quite difficult to solve such a nonconvex optimization problem. We will follow a similar two-stage iterative strategy in which the following two steps are repeated until the maximum number of iterations is reached: 1) subwindow assignment and 2) transmit power allocation and distance determination. It is first assumed that $p_k$ is set to $P_{\textrm{T}}/K$ and $d_k$ is set to the initial value $d_{\textrm{init}}$. To this end, in the first stage, subwindows are assigned to devices in order to maximize the TC. In the second stage, given the subwindow assignment, the transmit power and the distance of each device are determined so as to maximize the TC satisfying the minimum rate requirements.

In the following subsections, we present the proposed two-stage iterative allocation strategy precisely. Although the proposed strategy does not theoretically guarantee the optimal solution, we will empirically show that our subwindow assignment achieves almost the same performance on the TC as the optimal one via exhaustive search (see Section~\ref{SEC: Exhaustive_two_stage} for the numerical results). For the power allocation and the distance determination, we will also empirically show that the TC achieved by our iterative method is almost identical to the optimal one (see Fig.~\ref{Fig:TC-Distance-Absorption} in Section~\ref{SEC:PDD}).

\subsection{Subwindow Assignment}
In this stage, given that the values of $p_k$ and $d_k$ are fixed, we determine the optimal $\rho_{k,n}$ in terms of maximizing the TC by solving the following subwindow assignment problem:

\begin{align}\label{Prob-Subwindow}
\underset{\{ \rho_{k,n}\}}\max &~~ \sum_{k=1}^K \sum_{n=1}^N \rho_{k,n}d_k R_{k,n}(d_k)\nonumber\\
\textrm{s.t.}~&~~ \rho_{k,n} \in \{0,1\},~~~\forall k,n \nonumber\\
&~~ \sum_{k=1}^K \rho_{k,n} \leq 1,~~~\forall n \nonumber\\
&~~ \sum_{n=1}^N \rho_{k,n} = 1,~~~\forall k.
\end{align}

Similarly as in Section~\ref{SEC:Subwindow_fixed}, we apply the Hungarian method~\cite{Kuhn:55} to solve the problem above. Note that the minimum rate constraints do not need to be involved in the subwindow assignment stage since they can consistently be guaranteed in the next step by adjusting the power and the distance accordingly.

\subsection{Power Allocation and Distance Determination} \label{SEC:PDD}

In this subsection, we elaborate on the second stage of our allocation strategy, which allocates the transmit power to each device and determines the transmission distance of each device after carrying out the subwindow assignment. The power and distance determination problem is formulated as
\begin{align}\label{Prob-TC-Max}
\underset{\{p_k\},\{d_k\}}\max&~~ \sum_{k=1}^K d_k R_k \nonumber\\
\textrm{s.t.}~~
&~~ p_k \geq 0,~~~\forall k  \nonumber\\
&~~ d_k \geq 0,~~~\forall k  \nonumber\\
&~~ \sum_{k=1}^K  p_k \leq P_{\textrm{T}} \nonumber \\
&~~ R_k \geq R_{k,\textrm{th}},~~~\forall k.
\end{align}

Since the objective function $d_k R_k$ is highly nonconvex in $d_k$ although it is concave in $p_k$, it is not tractable to analytically (or numerically) find the optimal solution to the above problem in~\eqref{Prob-TC-Max}. The difficulty mostly comes from the exponential form of the molecular absorption loss in $R_k$ (i.e., the term $e^{-K_{\textrm{abs}}(f_n)d_k}$). In the following lemma, we first establish the optimality condition for distance--power pair $(d_k,p_k)$ of each device.

\begin{lemma}\label{Lemma:OptimalityCondition}
The TC of the $k$th device is a strictly quasiconcave function with respect to distance $d_k$. In addition, when there is no minimum rate requirement, the TC of the $k$th device is maximized if and only if the optimal distance--power pair $(d_k^o,p_k^o)$ satisfies the following optimality condition:
\begin{align}\label{OptimalityCondition}
\ln\left(1+\xi_k^o\right)\frac{1+\xi_k^o}{\xi_k^o}=2 + d_k^o K_{\textrm{abs}}(f_{n_k}),
\end{align}
where 
\begin{align} \label{EQ:SNR}
\xi_k^o=\frac{p_k^o G_t G_r}{\sigma^2}e^{-K_{\textrm{abs}}(f_{n_k})d_k^o}\left( \frac{c}{4\pi f_{n_k} d_k^o}\right)^2 
\end{align}
is the signal-to-noise ratio (SNR) of the $k$th device at the optimum and $n_k$ is the subwindow index assigned to the $k$th device.
\end{lemma}

\begin{IEEEproof}
See Appendix A.
\end{IEEEproof}

Now, using Lemma~\ref{Lemma:OptimalityCondition}, we would like to identify the following two fundamental operating regimes with respect to the minimum required rate $R_{k,\textrm{th}}$: the {\em TC-maximized} regime and the {\em distance-maximized} regime. 

\begin{proposition} \label{lemma:regimes}
The optimal distance $\bar{d}_k^o$ of the $k$th device with the minimum required rate $R_{k,\textrm{th}}$ is determined according to the operating regimes as follows:
\begin{align} \label{EQ:d_star}
  \bar{d}_k^o = \left\{
  \begin{array}{cl}
    \!\!d_k^o & \!\!\textrm{if~} R_{k,\textrm{th}} \leq W\eta_k^o~\textrm{(TC-maximized regime)}\\
   \!\! d_{k,\max} & \!\! \textrm{if~} R_{k,\textrm{th}} > W\eta_k^o~\textrm{(distance-maximized regime)},
  \end{array}
  \right.
\end{align}
where $\eta_k^o=\log_2(1+\xi_k^o)$ represents the spectral efficiency (in bps/Hz) of the $k$th device at the optimum and $d_{k,\max}$ is the maximum distance of the $k$th device satisfying
\begin{align}\label{Eq:BaseSNR-k}
  \frac{p_k G_t G_r}{\sigma^2}
  e^{-K_{\textrm{abs}}(f_{n_k})d_{k,\max}}
  \left( \frac{c}{4\pi f_{n_k} d_{k,\max}}\right)^2=2^{R_{k,\textrm{th}}/W}-1.
\end{align}
\end{proposition}
\begin{IEEEproof}
See Appendix B.
\end{IEEEproof}

\begin{figure}[!t]
  \centering
  \includegraphics[height=.37\textwidth]{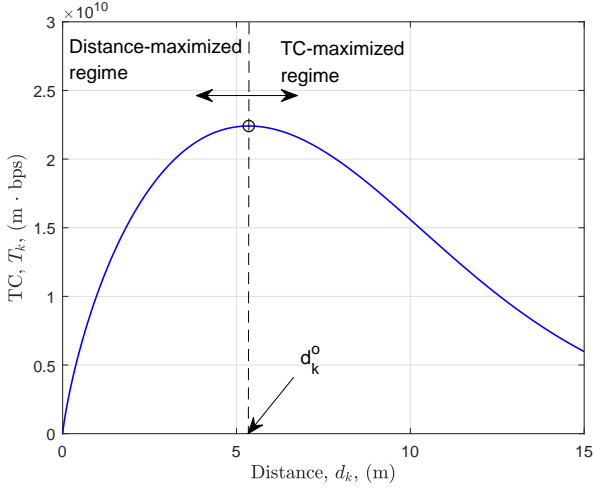}
  \caption{The TC versus the distance of a device where $f=500$ GHz, $W=1$ GHz, $G_t=G_r=15$ dBi, $K_{\textrm{abs}}(f)=0.2$, and $P_{\rm T}=10$ dBm. When $d_{k,\max}<d_k^o$, the device is in the distance-maximized regime while, when $d_{k,\max}\geq d_k^o$, the device is in the TC-maximized regime.}\label{Fig:TC-Distance}
\end{figure}

From Proposition~\ref{Lemma:OptimalityCondition}, our findings include that 1) in the TC-maximized regime, it is apparently not beneficial to increase the transmission distance over $d_k^o$ obtained in Lemma~\ref{Lemma:OptimalityCondition} in further improving the TC and 2) in the distance-maximized regime, the distance is determined as the maximum allowable distance such that the minimum required rate constraint is satisfied. In Fig.~\ref{Fig:TC-Distance}, the TC versus the distance (in $m$) is illustrated for $f=500$ GHz, $W=1$ GHz, $G_t=G_r=15$ dBi, $K_{\textrm{abs}}(f)=0.2$, and $P_{\rm T}=10$ dBm. As depicted in the figure, if $d_{k,\max}\geq d_k^o$, i.e., $R_{k,\textrm{th}} \leq W\eta_k^o$, then the regime is TC-maximized; otherwise, the regime is distance-maximized.

Next, let us explain how to discover the optimal distance $\bar{d}_k^o$ in~\eqref{EQ:d_star}. Although one can readily find $\bar{d}_k^o$ numerically via one-dimensional search when the power is given for each device, it would be a very difficult task to find the optimal distance--power pairs $(\bar{d}_k^o,\bar{p}_k^o)$ for all $k\in\{1,\ldots,K\}$ due to the fact that it may not be feasible to derive an analytical expression for the distance and the power. 

To tackle this challenge, we propose an iterative algorithm for solving the problem in \eqref{Prob-TC-Max}. More specifically, from~\eqref{EQ:SNR}, we use the following analytical expression for the power by fixing the molecular absorption loss with the term $d_k^{(i)}$ therein:
\begin{align}\label{Eq:power-iterative}
    p_k^{(i+1)} = \frac{\xi_k^{(i)}\sigma^2}{G_t G_r e^{-K_{\textrm{abs}}(f_{n_k})d_k^{(i)}}}\left(\frac{4\pi f_{n_k}d_k^{(i+1)}}{c}\right)^2,
\end{align}
where $d_k^{(i)}$, $p_k^{(i)}$, and $\xi_k^{(i)}$ indicate the distance, the power, and the SNR, respectively, of the $k$th device in the $i$th iteration. After finding the distance--power pair $(d_k^{(i+1)},p_k^{(i+1)})$ at the $(i+1)$th iteration, we update the molecular absorption loss by replacing $d_k^{(i)}$ with $d_k^{(i+1)}$. This process is repeated until convergence.
The following theorem establishes how to optimally determine the transmission distance for a fixed molecular absorption loss in~\eqref{Eq:power-iterative}.

\begin{theorem}
Suppose that the molecular absorption loss is $e^{-K_{\textrm{abs}}(f_{n_k})d_k^{(i)}}$ at the $(i+1)$th iteration for all $k=1,\ldots,K$. Then, the optimal distance of the $k$th device at the $(i+1)$th iteration, $\hat{d}_k^{(i+1)}$, is given by
\begin{align}\label{Optimal-distance}
  \hat{d}_k^{(i+1)} = \frac{G_t G_r e^{-K_{\textrm{abs}}(f_{n_k})d_k^{(i)}}\log_2\left(1+\xi_k^{(i)}\right)}{2 \nu^{(i+1)} \xi_k^{(i)} \sigma^2\left(4\pi f_{n_k}/c\right)^2},
\end{align}
where $\nu^{(i+1)}$ is determined in the sense of satisfying the following power constraint
\begin{align}
\sum_{k=1}^K \frac{\xi_k^{(i)}\sigma^2}{G_t G_r e^{-K_{\textrm{abs}}(f_{n_k})d_k^{(i)}}}\left(\frac{4\pi f_{n_k}\hat{d}_k^{(i+1)}}{c}\right)^2\leq P_{\textrm{T}}
\end{align}
and according to the operating regimes, $\xi_k^{(i)}$ is given by
\begin{align}  \label{xi_k}
  \xi_k^{(i)} = \left\{
  \begin{array}{cl}
    \tilde{\xi}_k^{(i)} & \textrm{if~} R_{k,\textrm{th}} \leq W\tilde{\eta}_k^{(i)}\\
    2^{R_{k,\textrm{th}}}-1 &  \textrm{if~} R_{k,\textrm{th}} > W\tilde{\eta}_k^{(i)}.
  \end{array}
  \right.
\end{align}
Here, $\tilde{\xi}_k^{(i)}$ and $\tilde{\eta}_k^{(i)}=\log_2(1+\tilde{\xi}_k^{(i)})$ are calculated for a given distance $d_k^{(i)}$ from the following equation:
\begin{align}
\ln\left(1+\tilde{\xi}_k^{(i)}\right)\frac{1+\tilde{\xi}_k^{(i)}}{\tilde{\xi}_k^{(i)}}=2 + d_k^{(i)} K_{\textrm{abs}}(f_{n_k}).
\end{align}
\end{theorem}
\begin{IEEEproof}
See Appendix C.
\end{IEEEproof}

Meanwhile, to avoid a possible oscillation between iterative solutions, we update the distance as
\begin{align}\label{Exponential-smoothing}
  d_k^{(i+1)} &= \alpha d_k^{(i)} + (1-\alpha) \hat{d}_k^{(i+1)},
\end{align}
where $\alpha>0$ is the smoothing factor for exponential smoothing. After obtaining $d_k^{(i+1)}$ from~\eqref{Exponential-smoothing}, we also update 
the power $p_k^{(i+1)}$ using~\eqref{Eq:power-iterative}. By repeatedly updating the distance $d_k^{(i+1)}$ and the power $p_k^{(i+1)}$ in this fashion, we finally yield the distance--power pair $(d_k^*,p_k^*)$ closely.

The pseudocode of the proposed iterative TC maximization method is described in Algorithm~\ref{Algorithm-Proposed}, where $m_{\textrm{out}}$ is the maximum number of iterations for the outer-loop and $\epsilon>0$ denotes the tolerance level determining the convergence speed. The effectiveness of the proposed algorithm is validated by a motivating example for $K=1$ as follows.

\begin{figure}[!t]
  \centering
  \includegraphics[height=.37\textwidth]{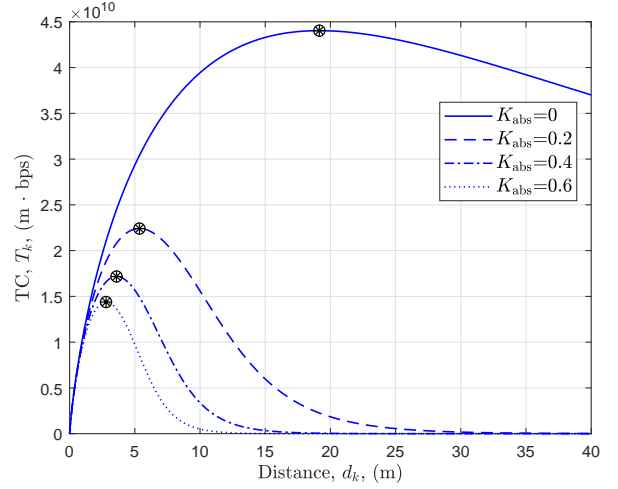}
  \caption{The TC of a device versus the distance for different absorption coefficients $K_{\textrm{abs}}(\cdot)\in\{0,0.2,0.4,0.6\}$, where $f=500$ GHz, $W=1$ GHz, $G_t=G_r=15$ dBi, $P_{\rm T}=10$ dBm, $\alpha=0.7$, and $\epsilon=10^{-6}$. Here, the circle (`o') and the asterisk (`*') denote the optimal and approximate solutions, respectively, to the TC maximization problem.}\label{Fig:TC-Distance-Absorption}
\end{figure}

\begin{example}
 In Fig.~\ref{Fig:TC-Distance-Absorption}, the TC of a device versus the transmission distance is plotted for different values of the absorption coefficient $K_{\textrm{abs}}(\cdot)$, where $f=500$ GHz, $W=1$ GHz, $G_t=G_r=15$ dBi, $P_{\rm T}=10$ dBm, $\alpha=0.7$, and $\epsilon=10^{-6}$. The optimal solutions, marked by the circle (`o'), are found via one-dimensional exhaustive search. The approximate solutions, marked by the asterisk (`*'), are also found via the above iterative allocation strategy. It is observed that the approximate solutions almost coincide with the optimal ones. 
\end{example}

In addition, we analyze the computational complexity of our proposed iterative TC maximization method (Algorithm 1) in the following remark.
\begin{remark}
The proposed method in Algorithm 1 consists of two stages: 1) the subwindow assignment (line 3) and 2) the power and distance determination (lines 7--13). For simplicity, it is assumed that $K=N$. For the subwindow assignment, we apply the Hungarian method whose complexity is given by $O(K^3)$~[57].\footnote{$f(x)=O(g(x))$ means the positive constants $M$ and $m$ exist such that $f(x)\le M g(x)$ for all $x>m$.} For the power and distance determination, the complexity is given by $O(m_{\textrm{in}} K)$, where $m_{\textrm{in}}$ is the number of iterations for the inner-loop (i.e., the while-loop). Hence, the overall complexity of Algorithm 1 is $O(m_{\textrm{out}}(K^3+m_{\textrm{in}} K))$, where $m_{\textrm{out}}$ is the number of iterations for the outer-loop.
\end{remark}

\begin{algorithm}[t!]
\caption{Proposed Iterative TC Maximization Algorithm}\label{Algorithm-Proposed}
\begin{algorithmic}[1]
\State Initialization: Set $p_1(0)=p_2(0)=\cdots=p_K(0)=P_{\textrm{T}}/K$, $d_k(0)=d_{\textrm{init}}~\forall k$, and $T_{\max}=0$. The TC of the $k$th devices at the $i$th iteration is denoted by $T_k^{(i)}$.
\For {$j\in [1:m_{\textrm{out}}]$} 
\State \multiline{Determine $\rho_{k,n}(j)~\forall k, n$ at the $j$th iteration by solving the problem in~\eqref{Prob-Subwindow} using the Hungarian algorithm for a given $d_k(j-1)$ and $p_k(j-1)$.}
\State $i\gets 0$.
\State $d_k^{(i)}\gets d_k(j-1)~\forall k$.
\State $p_k^{(i)}\gets p_k(j-1)~\forall k$.
\While {$\left|\sum_{k=1}^K T_k^{(i+1)}- \sum_{k=1}^K T_k^{(i)}\right|>\epsilon$}
\State $i\gets i+1$.
\State \multiline{Calculate $K_{\textrm{abs}}(f_{n_k})d_k^{(i-1)}$.}
\State \multiline{Calculate $\xi_k^{(i-1)}$ according to \eqref{xi_k}.}
\State \multiline{Determine $d_k^{(i)}$ and $p_k^{(i)}$ using \eqref{Optimal-distance}--\eqref{Exponential-smoothing} and \eqref{Eq:power-iterative}, respectively.}
\State Calculate $T_k^{(i)}$ for $k=1,\ldots,K$.
\EndWhile
\If {$T^{(i)}>T_{\max}$}
\State $T_{\max}\gets T^{(i)}$.
\EndIf
\State $d_k(j)\gets d_k^{(i)}~\forall k$.
\State $p_k(j)\gets p_k^{(i)}~\forall k$.
\EndFor
\end{algorithmic}
\end{algorithm}

\section{Numerical Evaluations and Discussions}~\label{SEC:NumericalResults}
In this section, we present numerical evaluations via intensive simulations to demonstrate the superiority of the proposed adaptive resource allocation method over benchmark methods in our multi-device Tera-IoT network. 

\subsection{Simulation Environments}

In our simulations, we adopt the THz band ranging from $0.5$ THz to $0.6$ THz since there is a large absorption loss at around 550 GHz that reflects the peculiarity of the THz band. The bandwidth of each subwindow, $W$, is set to $1$ GHz (unless otherwise stated), which enables us to accommodate up to 100 devices. Note that the system bandwidth is 100 GHz. The noise power spectral density, $N_0$, is set to $-168$ dBm/Hz (possibly including the interference power). The antenna gains $G_t$ and $G_r$ are 15 dBi each. In our proposed algorithm in Algorithm~\ref{Algorithm-Proposed}, it is assumed that $\alpha=0.7$, $\epsilon=10^{-6}$, and the maximum number of iterations, $m$, and the initial transmission distance, $d_{\textrm{init}}$, are assumed to be 5 and 10~m, respectively. Other simulation parameters are appropriately set according to each evaluation, which will be specified in the following subsections.

\subsection{TC with Fixed Distances}

In this subsection, when the distance between the AP and each device is given and fixed, we validate the impact and benefits of our TC maximization method in Section~\ref{SEC:OptimizationFixedDistances} in comparison with the sum rate maximization method as one of the most popular approaches in traditional resource allocation. Within a circular cell, the locations of $K$ devices are randomly generated for each simulation. This simulation is repeated 100 times. 

\begin{figure}[!t]
  \centering
  \subfigure[$K=80$.]{
  \includegraphics[height=.37\textwidth]{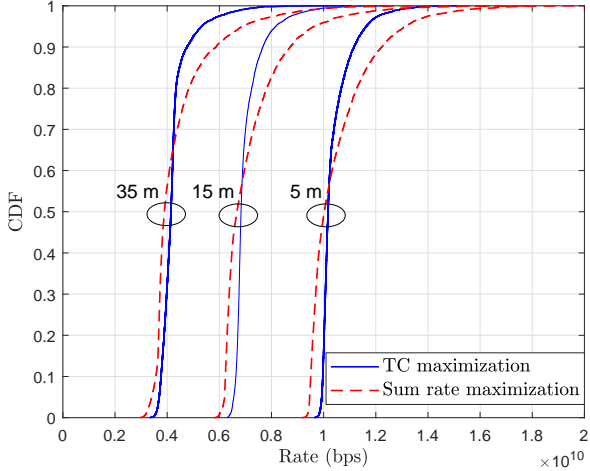} \label{Fig:CDF-Rate-K-80}}
  \centering
  \subfigure[$K=100$.]{
  \includegraphics[height=.37\textwidth]{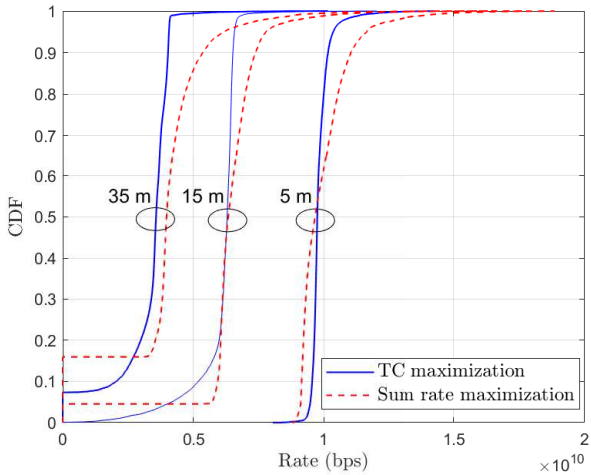} \label{Fig:CDF-Rate-K-100}}
  \caption{The CDFs of the transmission rate of each device for $N=100$ when the cell radius is given by 5, 15, and 35 m: (a) $K$=80; (b) $K$=100.}
  \label{Fig:CDF-Rate-K}
\end{figure}

In Fig.~\ref{Fig:CDF-Rate-K}, the cumulative density functions (CDFs) of the transmission rate of each device are plotted for $N=100$, $W=1$ GHz, and $P_{\textrm{T}}=40$ dBm. From Fig.~\ref{Fig:CDF-Rate-K-80}, it is seen that, when $K=80$, 1) at the 90-percentile point of the CDF, the transmission rate of the sum rate maximization method is much larger than that of the TC maximization method and 2) vice versa at the 5-percentile point. Since 
only some of 100 subwindows are assigned to 80 devices due to the fact that $K<N$, the subwindows around 550 GHz in which the molecular absorption loss is severe are not allocated at all (refer to Fig.~\ref{Fig:Pathloss-frequency}). However, from Fig.~\ref{Fig:CDF-Rate-K-100}, it is observed that, when $K=100$, the probability that a device has a zero transmission rate in the sum rate maximization method is largely increased as the cell radius increases. This is because the subwindows around 550 GHz are also allocated to some of devices whose transmission rate would be zero due to the very high path loss. On the other hand, in the TC maximization method, a non-zero transmission rate can still be achieved when the cell radius is smaller than 15 m. This implies that all devices are fairly served with non-zero transmission rates by using the TC maximization method.

In the following subsections, unless otherwise stated, we comprehensively demonstrate the superiority of our TC maximization method in Section~\ref{SEC:Optimization} for the case where the transmission distance between the AP and each device is allowed to vary.

\subsection{Benchmark Methods}

In this subsection, in comparison with the proposed adaptive TC maximization method with variable distances between the AP and each device, we present the following two benchmark methods: i) the distance maximization method and ii) the non-adaptive method. We begin by describing the distance maximization method.

We modify the distance maximization-based resource allocation strategies in~\cite{HanAkyildiz:TTST16,HanBicenAkyildiz:TSP16} so that such a modification is suitable for numerical evaluations in our current setting. In our benchmark method, the subwindow, the transmit power, and the transmission distance are jointly allocated to maximize the sum distance over all devices under the minimum rate constraints as below. For the subwindow assignment, the algorithm starts with the set of devices,  $\mathcal{U}=\{1,\ldots,K\}$, and the set of subwindows, $\mathcal{S}=\{1,\ldots,N\}$. We find the subwindow whose absorption coefficient $K_{\textrm{abs}}(\cdot)$ is the smallest, and then allocate this subwindow to the device that has the smallest required rate. After the allocated device and its subwindow are removed from $\mathcal{U}$ and $\mathcal{S}$, respectively, this assignment is repeated until all the devices are allocated to their subwindows. For the sum distance maximization, we formulate the power and distance allocation problem as
\begin{align}\label{Prob-Distance-Max}
\underset{\{p_k\},\{d_k\}}\max&~~ \sum_{k=1}^K d_k\nonumber\\
\textrm{s.t.}
&~~ p_k \geq 0,~~~\forall k  \nonumber\\
&~~ d_k \geq 0,~~~\forall k  \nonumber\\
&~~ \sum_{k=1}^K  p_k \leq P_{\textrm{T}} \nonumber \\
&~~ R_k \geq R_{k,\textrm{th}},~~~\forall k.
\end{align}

Since the above problem is a convex optimization problem, one can find the optimal solution using the Karush-Kuhn-Tucker (KKT) conditions. From the KKT conditions, the distance $d_k$ can be found in the sense of satisfying the following equation:
\begin{align}\label{Opt-Distance-Max-Distance}
  \nu\bar{\xi}_k \frac{\sigma^2(2d_k+ K_{\textrm{abs}}(f_{n_k})d_k^2)}{G_t G_r e^{-K_{\textrm{abs}}(f_{n_k})d_k}}
  \left(\frac{4\pi f_{n_k}}{c}\right)^2=1,
\end{align}
where $\bar{\xi}_k=2^{R_{k,\textrm{th}}}-1$, corresponding to the required SNR of the $k$th device, and $\nu$ is numerically determined to fulfill the total power constraint. From the distance $d_k$, the power $p_k$ can be obtained as
\begin{align}\label{Opt-Distance-Max-Power}
    p_k &=\frac{\bar{\xi}_k\sigma^2}{G_t G_r e^{-K_{\textrm{abs}}(f_{n_k})d_k}}\left(\frac{4\pi f_{n_k}d_k}{c}\right)^2.
\end{align}
The pseudocode of the distance maximization method is described in Algorithm~\ref{Algorithm-Distance-Max}. 

\begin{algorithm}[t!]
\caption{Distance Maximization Algorithm (Benchmark)}\label{Algorithm-Distance-Max}
\begin{algorithmic}[1]
\State Initialization: $\mathcal{U}=\{1,\ldots,K\}$ and $\mathcal{S}=\{1,\ldots,N\}$
\While {$\mathcal{U}\neq \phi$}
\State Find $k^*=\arg\min_{k} R_{k,\textrm{th}}$ and $n^*=\arg\min_n K_{\textrm{abs}}(f_n)$.
\State Assign the $n^*$th subwindow to the $k^*$th device.
\State $\mathcal{U}\gets \mathcal{U}\backslash k^*$ and $\mathcal{S} \gets \mathcal{S}\backslash n^*$.
\EndWhile
\State Determine the distance $d_k$ and the power $p_k$ for all $k$ using \eqref{Opt-Distance-Max-Distance} and \eqref{Opt-Distance-Max-Power}, respectively.
\end{algorithmic}
\end{algorithm}

Next, let us turn to describing the non-adaptive strategy. In order to assign subwindows to the devices, all the devices are first sorted in descending order of the required rates, and then the subwindows are assigned to the sorted devices from the first subwindow (i.e., from the lowest frequency). The molecular absorption loss is ignored in the subwindow assignment step. Afterwards, the transmission power is equally allocated to each device, and the distance is determined in terms of maximizing the TC while satisfying the rate requirements of all devices. In comparison with the proposed adaptive method, the impact of adaptive resource allocation (i.e., the joint power allocation and distance determination) on the TC maximization is demonstrated, which will be shown in later subsections. The pseudocode of the non-adaptive method is described in Algorithm~\ref{Algorithm-Non-Adaptive}.

\begin{algorithm}[t!]
\caption{Non-adaptive Algorithm (Benchmark)}\label{Algorithm-Non-Adaptive}
\begin{algorithmic}[1]
\State Initialization: Set $p_1=p_2=\cdots=p_K=P_{\textrm{T}}/K$, $n=1$, and $\mathcal{U}=\{1,\ldots,K\}$.
\While {$\mathcal{U}\neq \phi$}
\State Find $k^*=\arg\max_k R_{k,\textrm{th}}$.
\State Assign the $n$th subwindow to the $k^*$th device.
\State $\mathcal{U}\gets \mathcal{U}\backslash k^*$.
\State $n \gets n + 1$.
\EndWhile
\For {$k\in [1:K]$} 
\State Determine $d_k$ by solving \eqref{OptimalityCondition} with $p_k=P_{\textrm{T}}/K$.
\If {$R_k< R_{k,\textrm{th}}$}
\State Determine $d_k$ by solving \eqref{Eq:BaseSNR-k}.
\EndIf
\EndFor
\end{algorithmic}
\end{algorithm}

\subsection{TC with Respect to the Total Transmit Power}
\begin{figure}[!t]
  \centering
  \includegraphics[height=.37\textwidth]{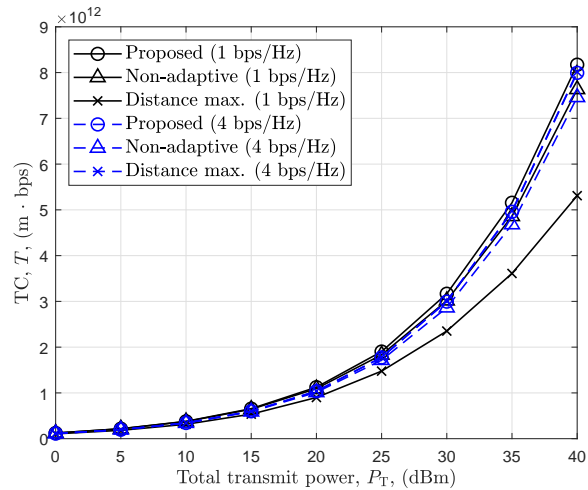}
  \caption{The TC versus the total transmit power $P_{\textrm{T}}$ for $K=N=100$, where $R_{k, \textrm{th}}/W \in \{1, 4\}$ bps/Hz for all $k$.} \label{Fig:TC-Power-same}
\end{figure}

In Fig. \ref{Fig:TC-Power-same}, the TC of three resource allocation strategies over the total transmit power $P_{\textrm{T}}$ is plotted for $K=N=100$, where two different rate constraints $R_{k, \textrm{th}}/W \in \{1, 4\}$ bps/Hz are assumed. It is observed that the TC values of the three methods are close to each other when $P_{\textrm{T}}$ is low since the limited transmit power becomes a bottleneck, but potential gains of the proposed method over two benchmark methods are clearly exhibited for high $P_{\textrm{T}}$. It can also be seen that the proposed method achieves a larger TC when $R_{k, \textrm{th}}/W = 1$ bps/Hz than its counterpart (i.e., the case of $R_{k, \textrm{th}}/W = 4$ bps/Hz). As stated in Lemma~\ref{lemma:regimes}, this is due to the fact that the TC is maximized when $R_{k, \textrm{th}}$ is small (i.e., the {\em TC-maximized} regime) while the distance is maximized when $R_{k, \textrm{th}}$ is large (i.e., the {\em distance-maximized} regime). In the following, the TC of the proposed adaptive method is more precisely compared with that of the two benchmark methods according to the total transmit power $P_{\textrm{T}}$.

\subsubsection{Comparison with the distance maximization method}
It is rather obvious that our adaptive TC maximization method is expected to be superior to the distance maximization method, which is inherently designed for enhancing the transmission distance. However, we aim at examining when the performance gap between these two methods is significant according to different heterogeneous rate requirements.
In the distance maximization method, the transmission rate is consistently determined in such a way that $R_k$ is set to  $R_{k,{\rm th}}$ in order to retain the minimum rate requirements quite tightly. As illustrated in Fig. \ref{Fig:TC-Power-same}, the TC of the distance maximization method is almost the same as that of the proposed method when $R_{k, {\rm th}}/W = 4$ bps/Hz. This is because the minimum required rate is large enough for all the devices to operate in the {\em distance-maximized} regime. On the other hand, when $R_{k, {\rm th}}/W = 1$ bps/Hz (i.e., the {\em TC-maximized} regime), the TC gap between the proposed and distance maximization methods becomes significant since, in the proposed method, the maximum TC can be achieved by reducing the transmission distance to the optimal distance, as addressed in Section~\ref{SEC:PDD}.

\subsubsection{Comparison with the non-adaptive method}
As shown in Fig. \ref{Fig:TC-Power-same}, the performance gap on the TC between the proposed adaptive and non-adaptive methods is non-negligible for the two rate constraints. In the non-adaptive method, not only the subwindows are assigned by ignoring the molecular absorption loss but also the transmit power is equally distributed to each device. Due to these non-adaptive factors, the performance of the non-adaptive method is inferior to the proposed iterative TC maximization method.

\begin{figure}[!t]
  \centering
  \includegraphics[height=.37\textwidth]{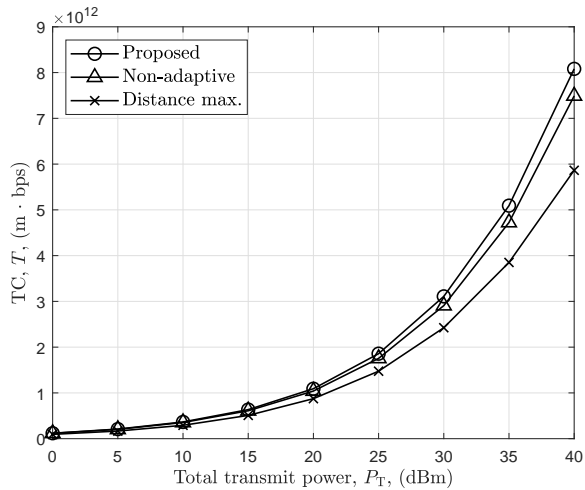}
  \caption{The TC versus the total transmit power $P_{\textrm{T}}$ for $K=N=100$ when $R_{k,\textrm{th}}/W=4$ bps/Hz for $k\in\{1,\cdots,50\}$ and $1$ bps/Hz for $k\in\{51,\cdots,100\}$.}\label{Fig:TC-Power-diff}
\end{figure}

\subsubsection{Heterogeneous rate requirements}
In Fig.~\ref{Fig:TC-Power-diff}, the TC values of the three resource allocation strategies over the total transmit power $P_{\textrm{T}}$ are plotted for $K=N=100$ under the {\em heterogeneous} rate requirements: $R_{k, \textrm{th}}/W = 1$ bps/Hz for $k\in\{1, \cdots, 50\}$ and $R_{k, \textrm{th}}/W = 4$ bps/Hz for $k\in\{51, \cdots, 100\}$. We recall that, when all the rate requirements are set to $R_{k, \textrm{th}}/W = 4$ bps/Hz, the TC of the proposed method is almost identical to that of the distance maximization method, as shown in Fig.~\ref{Fig:TC-Power-same}. However, under the heterogeneous rate requirement settings, the TC gap between the proposed and distance maximization methods becomes remarkably large since a larger TC value can be achieved by half of the devices whose required rates are set higher. Overall, if there are heterogeneous rate requirements including both $R_{k, \textrm{th}}/W \leq \eta_k^o$ and $R_{k, \textrm{th}}/W >\eta_k^o$, which may be feasible in realistic scenarios, then the superiority of the proposed method over the two benchmark methods can be more clearly revealed.

\begin{figure}[!t]
  \centering
  \includegraphics[height=.37\textwidth]{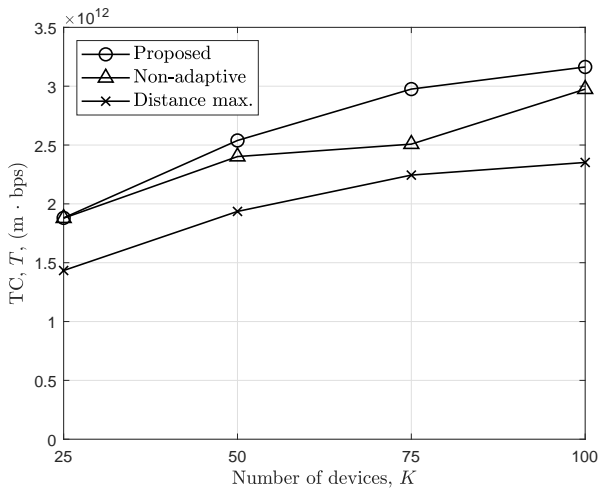}
  \caption{The TC versus the number of devices when $R_{k,\textrm{th}}/W=1$ bps/Hz for $k\in\{1,\cdots,25\}$; $2$ bps/Hz for $k\in\{26,\cdots,50\}$; $3$ bps/Hz for $k\in\{51,\cdots,75\}$; and $4$ bps/Hz for $k\in\{75,\cdots,100\}$ with $P_{\textrm{T}}=30$ dBm and $N=100$.}\label{Fig:TC-K}
\end{figure}

\subsection{TC with Respect to the Number of Devices}
In this subsection, as illustrated in Fig.~\ref{Fig:TC-K}, the TC of the three resource allocation strategies over the number of devices, $K$, is evaluated under heterogeneous rate requirements where $R_{k,\textrm{th}}/W=1$ bps/Hz for $k\in\{1,\cdots,25\}$; $2$ bps/Hz for $k\in\{26,\cdots,50\}$; $3$ bps/Hz for $k\in\{51,\cdots,75\}$; and $4$ bps/Hz for $k\in\{76,\cdots,100\}$. For example, when $K=25$, it follows that $R_{k,\textrm{th}}/W=1$ bps/Hz  for all $k$. It is assumed that $P_{\textrm{T}}=30$ dBm and $N=100$. From the figure, the following insightful observations are made: 1) the TC of the proposed adaptive method is almost identical to that of the non-adaptive method for $K=25$ due to the fact that both methods use the lowest subwindow frequency $f_n$ for $n=1,\ldots,25$ where the molecular absorption loss is not varying remarkably over subwindows and the variation of the spreading loss across subwindows is also quite small; 2) on the other hand, the performance gap between the proposed and distance maximization methods is very large for all values of $K$ since there are many devices whose minimum required rates are relatively small (i.e., $R_{k, \textrm{th}}\le W\eta_k^o$). Thus, the proposed method offers a significant gain especially when the number of devices constrained to heterogeneous rate requirements is large. In addition, it is seen that the performance gain of the non-adaptive method from $K=50$ to $K=75$ is negligible; this is because the molecular absorption loss is severe over frequency bands $f_n$ for $n=50,\ldots,75$ and the non-adaptive method still uses the equal power allocation.

This implies that the use of our adaptive resource allocation method in the THz band would be much more beneficial to dense network environments deploying a number of devices.

\begin{figure}[!t]
  \centering
  \includegraphics[height=.37\textwidth]{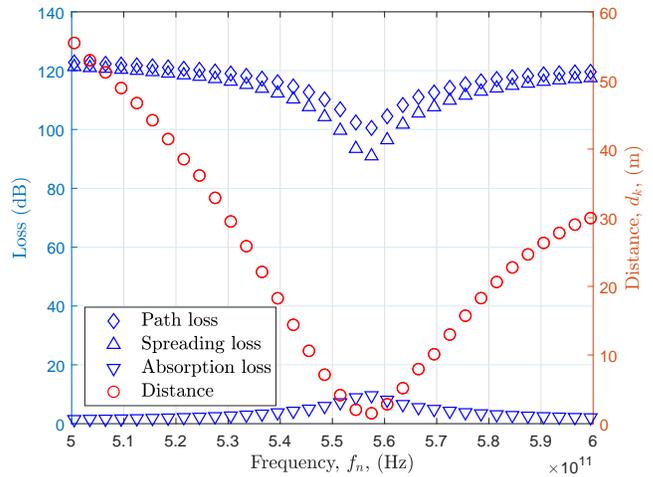}
  \caption{The losses and distance versus the frequency for the proposed adaptive method, where the left $y$-axis represents the path loss (without antenna gains), the spreading loss, and the molecular absorption loss, and the right $y$-axis represents the transmission distance of a device. Here, $K=N=100$, $P_{\textrm{T}}=40$ dBm, and $R_{k,\textrm{th}}=1$ bps/Hz for all $k\in\{1,\cdots,100\}$.}\label{Fig:LossDistance-Frequency}
\end{figure}

\subsection{Loss and Distance with Respect to the Frequency}
In the THz band, there exist several frequency bands in which the molecular absorption loss is very severe when the transmission distance is long (refer to Fig.~\ref{Fig:Pathloss-frequency}). Thus, it is highly crucial to allocate the resources precisely for such frequency bands incurring the high absorption loss so as to guarantee the maximum TC performance. For example, one of the frequency bands suffering from a high absorption loss is around 555 GHz; hence, the transmission distance of a device using the frequency around 555 GHz should be very short to avoid the very high absorption loss. In Fig.~\ref{Fig:LossDistance-Frequency}, for the proposed adaptive method, three types of losses, including the path loss (without antenna gains), the spreading loss, and the absorption loss, as well as the transmission distance are plotted with respect to the frequency ranging from 500 GHz to 600 GHz when $K=N=100$, $P_{\textrm{T}} = 40$ dBm, and $R_{k,\textrm{th}}=1$ bps/Hz for all $k\in\{1,\cdots,100\}$. As expected above, the distance allocated by the proposed adaptive method for the device(s) using the frequency around 555 GHz tends to be very short due to a huge amount of absorption loss. 
These results demonstrate that our proposed method is capable of allocating the limited resources to the devices by judiciously taking both the spreading loss and the molecular absorption loss into account.

\subsection{Rate--Distance Trade-off}

\begin{figure}[!t]
  \centering
  \includegraphics[height=.37\textwidth]{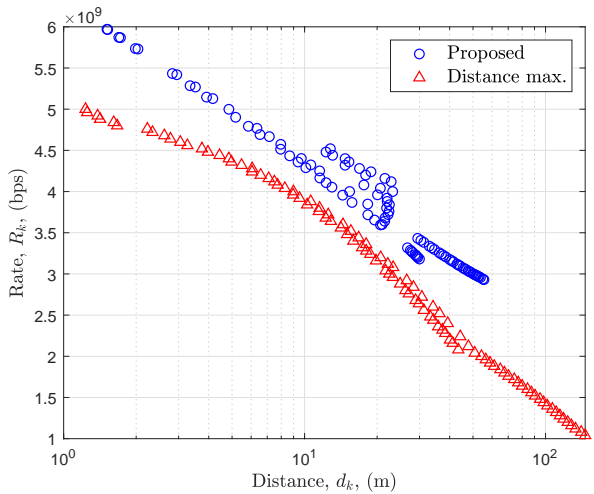}
  \caption{The allocated rate versus the determined transmission distance for $K=N=100$ and $P_{\textrm{T}}=40$ dBm, where the proposed and distance maximization schemes are employed as $R_{k, {\rm th}}/W =1.04+0.04(k-1)$ bps/Hz for $k\in\{1,\cdots,100\}$.}\label{Fig:Rate-Distance}
\end{figure}

In this subsection, we numerically characterize the fundamental trade-off between the rate and the transmission distance for the proposed and distance maximization methods. In Fig.~\ref{Fig:Rate-Distance}, the rate of each device versus the distance is plotted for $K=N=100$ and $P_{\textrm{T}}=40$ dBm under heterogeneous rate requirements where $R_{k, {\rm th}}/W =1.04+0.04(k-1)$ bps/Hz for $k\in\{1,\cdots,100\}$. It is observed that the proposed adaptive method achieves higher transmission rates than those of the distance maximization method at the same distance. In the distance maximization method, a lower transmit power is allocated to the devices constrained to the high rate requirements and the remaining transmit power is exploited to increase the transmission distance of the devices having the low rate requirements (see Fig.~\ref{Fig:Rate-Distance}). In other words, in the distance maximization method, the transmit power is primarily used to increase the distance, whereas the transmission rates are set tightly to the minimum required rates. On the other hand, in the proposed adaptive method, when the minimum required rates are low (i.e., $R_{k, {\rm th}}\le W\eta_k^o$), the transmission rates are increased up to $W\eta_k^o$ while the transmission distance of the devices constrained to the low rate requirements is sacrificed. As depicted in Fig.~\ref{Fig:Rate-Distance}, this implies that, due to the heterogeneous rate requirements, there can be possibly multiple devices whose transmission rates are different from each other at the same transmission distance from the AP. These results exhibit that there is a net improvement in the fundamental rate--distance trade-off achieved by the proposed TC maximization method over the distance maximization method.

\subsection{Exhaustive Subwindow Assignment} \label{SEC: Exhaustive_two_stage}

\begin{figure}[!t]
  \centering
  \includegraphics[height=.37\textwidth]{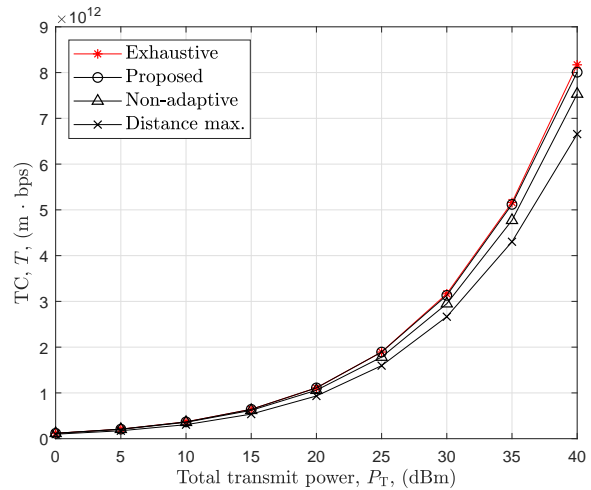}
  \caption{The TC versus the total transmit power $P_{\textrm{T}}$ for $K=N=5$, where $R_{k,\textrm{th}}/W=k$ bps/Hz for $k\in\{1,\ldots,5\}$ and $W=20$ GHz.}\label{Fig:TC-Exhaustive}
\end{figure}

In order to validate the effectiveness of our proposed two-stage strategy in which the subwindows are assigned in the first stage and the distance and the power are allocated in the second stage, we evaluate the performance of the optimal method via exhaustive search with respect to the subwindow assignment. More specifically, for all possible subwindow assignments, the TC is computed by determining the distance and the power as described in Algorithm~\ref{Algorithm-Proposed}. In Fig.~\ref{Fig:TC-Exhaustive}, the TC of four resource allocation methods including the above optimal one over the total transmit power $P_{\textrm{T}}$ is plotted for $K=N=5$, where $R_{k,\textrm{th}}=k$ (bps/Hz) for $k\in\{1,\ldots,5\}$ and $W=20$ GHz.\footnote{Note that $N$ is set to a small value so that simulations are completed within a reasonable time. The computational complexity of the exhaustive search-based method will be prohibitively high for large $N$.} One can see that the performance of the proposed two-stage method is quite close to (essentially the same as) that of the exhaustive search. Hence, as long as the subwindow assignment is concerned, our method is capable of achieving the near optimal performance.


\section{Concluding Remarks}~\label{SEC:Conclusion}
In this paper, we presented a new joint resource allocation strategy in the sense of maximizing the TC as a KPI in the dense Tera-IoT network having a number of small-scale devices. More specifically, we formulated two optimization problems and solved them while taking into account both the spreading and molecular absorption losses judiciously. First, when the transmission distance to each device is fixed, we presented how to effectively solve the joint subwindow and power optimization problem. Second, when the transmission distance between the AP and each device varies, we formulated our optimization such that the subwindow, transmit power, and distance are jointly optimized under the minimum heterogeneous rate constraints, and then proposed an effective two-stage strategy that iteratively performs subwindow assignment and power--distance determination. We also investigated how the optimal transmission distance of each device is determined according to the two fundamental operating regimes, including the TC-maximized and distance-maximized regimes, with respect to the rate constraint. Through comprehensive evaluations via simulations, we validated the superiority of the proposed TC maximization method. For the TC maximization with fixed distances, unlike the sum rate maximization method, our method was shown to fairly serve all devices with non-zero transmission rates. For the TC maximization with variable distances, we demonstrated the gain of the proposed adaptive method over two benchmark methods in terms of the TC. Furthermore, our results verified that 1) the proposed method is much beneficial to the case where the number of devices constrained to heterogeneous rate requirements is large and 2) the overall rate--distance trade-off can be improved by employing the proposed TC maximization strategy, which indicates that a higher rate can be achieved for a given transmission distance. 

Potential avenues of future research in this area include resource allocation that maximizes the TC by taking into account not only the multiple APs but also NOMA systems. The bandwidth allocation among multiple users would also be an interesting future topic.

\appendices

\section{Proof of Lemma 1}
We start by proving that the rate--distance product term $T_k$, corresponding to the TC of device $k$, is a strictly quasiconcave function over the distance $d_k\ge 0$. The first derivative of $T_k$ with respect to $d_k$ is given by
\begin{align}\label{TC-first-derivative}
\frac{d T_k}{d d_k} &= W\log_2 \left(1+\frac{p_k G_t G_r e^{-K_{\textrm{abs}}(f_{n_k})d_k}}{\sigma^2} \left( \frac{c}{4\pi f_{n_k} d_k}   \right)^2 \right)
\nonumber\\
&~~~- \frac{W(2+d_k K_{\textrm{abs}}(f_{n_k}))}{\ln 2 \left( 1+ \frac{\sigma^2}{p_k G_t G_r e^{-K_{\textrm{abs}}(f_{n_k})d_k}} \left( \frac{4\pi f_{n_k} d_k}{c}   \right)^2 \right)}
\nonumber\\
&=W\log_2 (1+\xi_k) - \frac{W(2+d_k K_{\textrm{abs}}(f_{n_k}))}{\ln 2}\frac{1}{1+1/\xi_k},
\end{align}
where $\xi_k=\frac{p_k G_t G_r e^{-K_{\textrm{abs}}(f_{n_k})d_k}}{\sigma^2} \left( \frac{c}{4\pi f_{n_k} d_k}\right)^2$. After multiplying both sides of \eqref{TC-first-derivative} by $(1+1/\xi_k)/W$, we have
\begin{align}\label{TC-first-derivative-mod-1}
    &\frac{1}{W}\left(1+\frac{1}{\xi_k}\right)\frac{d T_k}{d d_k} 
    \nonumber\\
    &=\left(1+\frac{1}{\xi_k}\right)\log_2 (1+\xi_k) - \frac{2+d_k K_{\textrm{abs}}(f_{n_k})}{\ln 2}.
\end{align}

The first term in the right-hand side (RHS) of \eqref{TC-first-derivative-mod-1} is increasing in $\xi_k$, or equivalently, decreasing in $d_k$. The second term in the RHS of \eqref{TC-first-derivative-mod-1} is also decreasing in $d_k$. When $d_k$ tends to zero, \eqref{TC-first-derivative-mod-1} approaches
\begin{align}
    \log_2 (1+\xi_k) - \frac{2}{\ln 2}, \nonumber
\end{align}
which is positive. Thus, there is a point $d_k^o$ such that \eqref{TC-first-derivative-mod-1} is positive for $d_k<d_k^o$ and \eqref{TC-first-derivative-mod-1} is negative for $d_k>d_k^o$. In other words, there is a point $d_k^o$ such that $T_k$ is strictly increasing for $d_k<d_k^o$ and $T_k$ is strictly decreasing for $d_k>d_k^o$. Therefore, $T_k$ is a strictly quasiconcave function over $d_k$. 

In general, a local optimum is not necessarily the global optimum for quasiconcave functions. For strictly quasiconcave functions, however, a local optimum is also the global and unique optimum. By setting the first derivative in~\eqref{TC-first-derivative} to zero, the optimality condition on $d_k$ can be obtained as
\begin{align}
  \ln (1+\xi_k)\left(1+\frac{1}{\xi_k}\right) 
  = 2+d_k K_{\textrm{abs}}(f_{n_k}),
\end{align}
which thereby comes to the conclusion that the condition on the optimal $(d_k^o,p_k^o)$ is finally given by (\ref{OptimalityCondition}).
This completes the proof of this lemma.

\section{Proof of Proposition 1}
If $R_{k,\textrm{th}} \leq W\eta_k^o$ (i.e., the minimum required rate is set low), then the optimal pair of $(d_k^o,p_k^o)$ in~\eqref{OptimalityCondition} can satisfy the minimum rate requirement while enabling us to maximize the rate--distance product of the corresponding device $k$. Thus, the optimal distance under the rate requirement, $\bar{d}_k^o$, is the same as $d_k^o$. In this case, the device $k$ is fundamentally in the TC-maximized regime. 

On the other hand, if $R_{k,\textrm{th}} > W\eta_k^o$ (i.e., the minimum required rate is set high), then the optimality condition in~\eqref{OptimalityCondition} no longer holds. For $R_{k,\textrm{th}} > W\eta_k^o$, therefore, the pair of $(d_k,p_k)$ should be determined in the sense of increasing the transmission distance as large as possible while satisfying the minimum required rate constraint $R_{k,\textrm{th}}$. Thus, the optimal distance under the rate requirement, $\bar{d}_k^o$, is given by $d_{k,\max}$, which is the solution to \eqref{Eq:BaseSNR-k}. In this case, the corresponding device is fundamentally in the distance-maximized regime. This completes the proof of this proposition.

\section{Proof of Theorem 1}
By substituting \eqref{Eq:power-iterative} into \eqref{Prob-TC-Max}, we can reformulate the power and distance determination problem as the following iterative distance determination problem:
\begin{subequations} \label{Pr:re-distance}
\begin{align}
\underset{\{\hat{d}_k^{(i+1)}\}}\max &~~ \sum_{k=1}^K  \hat{d}_k^{(i+1)} W \log_2 \left(1+\xi_k^{(i)}\right)
\label{Pr:re-distance-1}\\
\textrm{s.t.}&~~ \hat{d}_k^{(i+1)} \geq 0,~~~\forall k \label{Pr:re-distance-2}\\
&~~ \sum_{k=1}^K \frac{\xi_k^{(i)}\sigma^2}{G_t G_r e^{-K_{\textrm{abs}}(f_{n_k})d_k^{(i)}}}\left(\frac{4\pi f_{n_k}\hat{d}_k^{(i+1)}}{c}\right)^2\leq P_{\textrm{T}}, \label{Pr:re-distance-3}
\end{align}
\end{subequations}

Since the problem in~\eqref{Pr:re-distance} becomes convex, it is possible to find the optimal solution using the KKT conditions, which are given by
\begin{align}\label{KKT}
    &-\hat{d}_k^{(i+1)}\leq 0,~~~\forall k \nonumber\\
    &\sum_{k=1}^K \frac{\xi_k^{(i)}\sigma^2}{G_t G_r e^{-K_{\textrm{abs}}(f_{n_k})d_k^{(i)}}}\left(\frac{4\pi f_{n_k}\hat{d}_k^{(i+1)}}{c}\right)^2 - P_{\textrm{T}}\leq 0\nonumber\\
    &\lambda_k^{(i+1)}\geq 0,~~~\forall k\nonumber\\
    &\lambda_k^{(i+1)} \hat{d}_k^{(i+1)} = 0,~~~\forall k\nonumber\\
    &\nu^{(i+1)}\left(\sum_{k=1}^K \frac{\xi_k^{(i)}\sigma^2}{G_t G_r e^{-K_{\textrm{abs}}(f_{n_k})d_k^{(i)}}}\left(\frac{4\pi f_{n_k}\hat{d}_k^{(i+1)}}{c}\right)^2 \hspace{-5pt}- P_{\textrm{T}}\hspace{-4pt}\right)\hspace{-3pt}=\hspace{-2pt}0\nonumber\\
    &-\log_2\left(1+\xi_k^{(i)}\right)-\lambda_k\nonumber\\
    &+\frac{2\nu\xi_k^{(i)}\sigma^2}{G_t G_r e^{-K_{\textrm{abs}}(f_{n_k})d_k^{(i)}}}\left(\frac{4\pi f_{n_k}\hat{d}_k^{(i+1)}}{c}\right)\frac{4\pi f_{n_k}}{c} = 0,~~~\forall k,
\end{align}
where $\lambda_k^{(i)}$'s for $k=1,\ldots,K$ and $\nu^{(i)}$ are the Lagrangian multipliers associated with the inequalities in~\eqref{Pr:re-distance-2} and~\eqref{Pr:re-distance-3} in the $i$th iteration, respectively. From the KKT conditions in~\eqref{KKT}, $\hat{d}_k^{(i+1)}$ is finally given by (\ref{Optimal-distance}). This completes the proof of this theorem.



\ifCLASSOPTIONcaptionsoff
  \newpage
\fi



\bibliographystyle{IEEEtran}
\bibliography{IEEEabrv,references}

\end{document}